\documentclass{sig-alternate-05-2015}

\usepackage{amsmath}
\usepackage{booktabs}
\usepackage{array}
\usepackage{longtable}
\usepackage{makecell}
\usepackage{graphicx}
\usepackage{epstopdf}
\usepackage{float}
\usepackage{multirow}
\usepackage{balance}
\usepackage{bm}
\usepackage{algorithm}
\usepackage{algorithmic}
\usepackage{subfigure}
\usepackage{makecell}

\makeatletter
\def\@copyrightspace{\relax}
\makeatother

\begin{document}

\title{Learning Sentimental Influences from Users' Behaviors}

\numberofauthors{5} 
\author{
Shenghua Liu,$^{1}$ Houdong Zheng,$^{2}$ Huawei Shen,$^{1}$ Xiangwen Liao,$^{2}$ Xueqi Cheng$^{1}$\\ 
\normalsize{$^{1}$CAS Key Laboratory of Network Data Science \& Technology}\\
\normalsize{Institute of Computing Technology, Chinese Academy of Sciences}\\
\normalsize{$^{2}$Mathematics and Computer Science School, Fuzhou University}\\
\normalsize{liushenghua@ict.ac.cn, shenghuawei@ict.ac.cn, liaoxw@fzu.edu.cn, cxq@ict.ac.cn}
       }

\maketitle

\begin{abstract}
    Modeling interpersonal influence on 
different sentimental polarities is a fundamental
problem in opinion formation and viral marketing. 
There has not been seen an effective solution for learning
sentimental influences from users' behaviors yet.
Previous related works on information propagation 
directly define interpersonal
influence between each pair of users as a parameter,
which is independent from each others, even if the influences
come from or affect the same user.
And influences are learned from user's propagation
behaviors, namely temporal cascades, while
sentiments are not associated with them.
Thus we propose to model the interpersonal influence
by latent influence and susceptibility matrices defined
on individual users and sentiment polarities.
Such low-dimensional and distributed representations 
naturally make
the interpersonal influences related to the same
user coupled with each other, and in turn, reduce the model complexity.
Sentiments act on different rows of parameter matrices, depicting
their effects in modeling cascades.
With the iterative optimization algorithm of projected stochastic gradient descent 
over shuffled mini-batches and Adadelta update rule,
negative cases are repeatedly sampled with the distribution
of infection frequencies users, for reducing computation cost and
optimization imbalance.
Experiments are conducted on Microblog dataset. The results
show that our model achieves better performance
than the state-of-the-art and pair-wise models.
Besides, analyzing the distribution of learned users' sentimental 
influences and susceptibilities results some interesting discoveries.

\end{abstract}


\section{Introduction}
\label{secintro}
Collective opinions concisely form by a repeated process that 
a user who sees and agrees with a sentimental content, forwards, shares,
or ``Likes'' to feed her virtual community,
resulting in a temporal cascade of users' behaviors.
As such, users who have taken actions become infective to
others in their communities, encouraging their communities
to a certain extend to take the same action by 
interpersonal influences.
Each pair of users has a specific influence, 
especially when there exits some kind
of social relation~\cite{goyal2010learning,aral2012identifying}.
Therefore, both opinion formation~\cite{watts2007influentials,bindel2015bad} 
and viral marketing~\cite{RichardsonKDD2002,leskovec2007dynamics} see
the importance of learning sentimental influences between 
users, 
with which one can better model the dynamics of cascades~\cite{goyal2010learning}, 
and maximize influence~\cite{kempe2003maximizing,gionis2013opinion,ChengCIKM2013}.

\begin{figure}[t]
\centering
\includegraphics[width=0.30\textwidth]{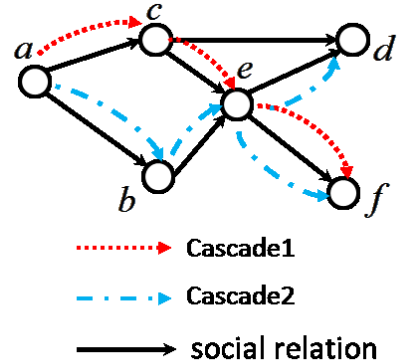}
\caption{Motivations underlying our model. Example of cascades to
illustrate the overfitting problem suffered by pair-wise models.}
\label{fig:motivation}
\end{figure}

Among the existing works,
it seldom sees an effective one for estimating sentimental
influence between pairs of users as far as we are concerned.
Most of studies focus on the process how users repeatedly 
update their opinions, the consensus they can reach
~\cite{degroot1974reaching,suchecki2005voter,bindel2015bad},
and opinion influence maximization \cite{gionis2013opinion},
assuming that the interpersonal influences as
edge weights are given or equally assigned.
In the related domain of information propagation, influences learning
has been studied, without consideration of different propagation 
behaviors on sentiments, although some studies modeled them
on the distribution of content topics \cite{Liu2010,Du2013}.
Moreover, 
Goyal et. al \cite{goyal2010learning} ``learned'' 
interpersonal influences 
by counting the successful propagation pairs. And with Bernoulli or Jaccard
Index model, they estimated the influences as propagation probabilities.
However, there usually record the times of users getting infected,
while such successful propagation pairs that user $v$ infects user $u$
are rarely observed, or hardly traced. 
It limits the application of such a method.
And NetInf \cite{goyal2010learning} used an exponential or 
power-law form of incubation time between 
a pair of infected users to estimate the interpersonal influence,
with empirically assigned parameters.
Afterward, more novel models were proposed 
to learn interpersonal influence by 
maximizing the likelihood of observed cascades,
which were proved to be more 
effective~\cite{saito08,GomezRodriguez2011,gomez2013structure}.
However, they used a free scalar parameter directly defined
on a pair of users to represent the interpersonal
influence. On one hand, the parameters are independent, even if
the influences are acted by or apply to the same user.
On the other, such a pair-wise parameter cannot 
be trained if there are no observations of propagations
between the user pair, $e.g.$ users $c$ and $d$ 
do not appear in the same cascade, infected
in Cascade1 and Cascade2 respectively, 
as Figure \ref{fig:motivation} shows.
In that case, even though $c$ and $d$ have a
relationship, and $c$, $e$ and $d$ form a social triangle,
a zero or some empirically small constant
is assigned, implying that it is never or
seldom to successfully propagate between them 
in the future. 
In another way, 
Aral and Walker \cite{aral2012identifying}
proposed to model the interpersonal influence
by engineered features and corresponding linear coefficients 
learned for individual users other than user pairs directly.
But users' properties may not be available or 
easy to extract in other applications.

To fill the blank of previous works, 
we thus propose to learn distributed representations
of users' influences and susceptibilities on sentiments. 
With such representations, we model the 
interpersonal influences and their decaying with the elapsed time
in the hazard function of survival model,
and maximize the likelihood of the observed the 
behaviors of users taking or not taking the actions 
in sentimental cascades. 
Hence, the interpersonal influences, between two pairs of users,
and with the same acting or applied user, can be coupled 
due to the corresponding representation of the same user. 
For example, 
the interpersonal influence of users $c$ to $d$, 
couples with that of users $c$ to $e$
by the corresponding representation defined on user $c$ 
as in Figure \ref{fig:motivation}.
Besides, it requires much fewer parameters, $i.e.$, $O(n)$
for $n$ users, instead of $O(n^2)$ parameters for user pairs,
beneficial to reducing the model complexity, 
and in turn combating the overfitting problem of assigning an empirical
propagation probability.
Moreover, the number of infected users in a cascade is usually much less than 
that of uninfected ones as negative cases. Considering all the negative
users prevents the model from being applied to a real large dataset and balancing the
optimization. 
Thus a negative sampling is employed to consider the expectation of negative
cases instead, emphasizing the frequently infected users in other cascades.
Finally a mini-batch Stochastic Gradient Decent (SGD) 
algorithm with Projected Gradient (PG)
is designed to learn the model, and Adadelta is used to adjust 
the learning rate adaptively. In such a scheme, negative sampling is
repeated in each iteration with a small number of samples each time,
to approximate the expectation. 

A set of cascades with different sentiments 
are collected from Microblog, covering a group of
users who interact at a frequent level.
Comparing with the state-of-the-art models, including
above Bernoulli and Jaccard Index estimation methods, 
and pair-wise models, our model
achieves better performances on the tasks of
predicting cascade dynamics, ``who will be retweeted'', and 
cascade size with users' representations.
Besides, it can be seen that learning influences separately
on different sentimental polarities, mostly benefits 
the performances on both tasks, even if more parameters are brought in.
As last, users' representations on different sentiments
are analyzed as well.
And we find that users may have different influences on
different sentiments, and are susceptible to different
polarities. The ``original influentials'' are creative to post
original attractive messages, while the ``secondary influentials'' gain their influence credits 
by hunting and advertising for interesting
messages that already exists in the system.

The rest of the paper is organized as follows.
Section \ref{secRelatedWorks} studies the existing and related works, 
and the motivation and our model are
described in section \ref{secmotmod}, which the parameter learning 
algorithm is given. At last, experiments
and result analysis are reported in section \ref{secexperiments}, and
section \ref{secconclusions} concludes the whole work.

\section{Related Work}
\label{secRelatedWorks}

Sentiment propagation and opinion formation have
attracted many research works.
\cite{zafarani2010sentiment,kramer2014experimental}
experimentally showed that users' sentiments were
influenced by that of others surrounding them on
LiveJournal dataset and Facebook dataset separately.
\cite{bae2012sentiment} used Granger causality analysis
to show that sentiment change of audiences were 
related to the landscape of popular users in Twitter.
As for modeling opinion dynamics, 
successful models were proposed, including
Sznajd model~\cite{sznajd2000opinion}, Deffuant model 
~\cite{deffuant2000mixing}, and  Hegselmann and Krause 
model ~\cite{hegselmann2002opinion},
which produced agreeing results.
Moreover, \cite{rodrigues2005surviving} 
extended Sznajd model to
complex networks. 
Deffuant et al.~\cite{deffuant2000mixing} modeled the 
process of opinion dynamics that 
randomly select two users, and change
their opinions to reduce the difference.
\cite{hestenes1969multiplier} modeled to change users' opinions
according to the arithmetic average of that of their
neighbors, and Fortunato et al.~\cite{fortunato2005vector}
extended the model with multi-dimensional opinion vector,
instead of a scalar opinion value.
Besides, Suchecki et al.~\cite{suchecki2005voter} studied 
Voter model in scale-free network, small-world network,
and random network. 
A recent work \cite{bindel2015bad} by Bindel et al. 
discovered that traditional models including DeGroot
model \cite{degroot1974reaching} finally converged
to a state of consensus under a set of general conditions,
while it is rare in real opinion dynamics. Hence it
proposed to model with users' intrinsic beliefs in 
a game theory, which counterbalanced the opinions at
Nash equilibrium.
In addition, Gionis et al.~\cite{gionis2013opinion}
studied the overall positive opinion maximization
problem, adopting the game model~\cite{bindel2015bad} 
of opinion dynamics.
\cite{de2015learning} modeled that a user's opinion 
was generated from her latent opinion distribution, 
based on self-excited Hawkes process influenced by her neighbors. 

The body of above works is mostly on opinion dynamics
and maximization, assuming the sentimental
influences between connected users were equal.
It does not confirm to our observations in real life,
in which a minority of influential users infect an
exceptional number of their peers~\cite{katz1955personal}, 
and there are a mass of easily influenced 
users~\cite{watts2007influentials}. 
Thus, as a fundamental problem, sentimental influences were
ever estimated by counting under Bernoulli assumption, or
a threshold rule~\cite{watts2007influentials}. 
As far as we know, an effective method of learning sentimental 
influences from users' behaviors remains unexplored.

Nevertheless, there were quite a few successful works
on estimating interpersonal influences in
the related domain of information propagation.
Some of them made efforts to extract features that 
are related to propagation probability 
and learned from the observed information cascades.
Crane et al.~\cite{crane2008robust} measured the response
function of information propagation dynamics in social systems with endogenous
and exogenous factors. 
Artzi et al.~\cite{Artzi2012} predicted whether a user would respond to or
retweet a message, $i.e.$ get influenced, by classifying with demographic and
content features. 
In a way other than feature extractions,
Tang et al. ~\cite{tang2009social} proposed 
topic factor graph (TFG) to model the generative process of 
the topic-level social influence on large networks, 
by finding a topic distribution for each user.
And \cite{Liu2010} proposed
a probabilistic factor graph to model the direct and 
indirect influences between adjacent and non-adjacent users
of heterogeneous network. 
Saito et al.~\cite{saito08} 
learned the propagation probability 
between neighbors of a directed network 
under independent cascade model, using the orders
of users getting influenced as training data.
And Goyal et. al \cite{goyal2010learning} proposed to estimate 
the interpersonal influences 
in a counting manner, with assumptions of Bernoulli model 
and Jaccard Index separately. They estimated the influences 
as propagation probabilities.
NetInf \cite{gomez2010inferring}
adopted both
exponential and a power-law incubation time models with 
fixed parameters as
pair-wise  probability to infer the underlying network.
Besides, there are also
a series of works learning a propagation probability between
any pair of users with survival model and its variants
to infer underlying networks with the transmission rates.
NetRate \cite{GomezRodriguez2011}
used survival theory to model transmission
rate between every pair of users, 
which was viewed as an edge weight for the pair. 
And \cite{gomez2013modeling} then
modeled the hazard rate in survival model with
additive and multiplicative risks separately to improve the performance of
cascade size prediction.
Afterward, InfoPath \cite{gomez2013structure} was proposed to
learn time-varying transmission rates
for user pairs as the edge weights of the hidden dynamic network. 
Taken together, these methods work in a pair-wise
manner, $i.e.$, 
they learned the propagation probability between pairs of users,
fundamentally different from the proposed method in this paper which focuses on
inferring user-specific influence and susceptibility from historical cascades.
The features in influence and susceptibility representations were analyzed
by \cite{aral2012identifying}, showing that propagation probability
were determined by the two feature vectors,
and learned the correlations
between users' attributes to identify influential or susceptible users. 
\cite{wang2015learning} then proposed a sequence model to learn a user's
latent representation of influence and susceptibility, based
on the orders of users' getting infected. 
In this work, we propose to learn the distributed representations
of a user on sentiments, and continuous time model is employed to
consider the infected times of users and the effect of the elapsed time 
on interpersonal influences, rather than their orders only.

\section{Learning Sentimental Influence}
\label{secmotmod}

A cascade is the snapshot of a propagation process, 
recording the times that users take actions on the same target,
such as a piece of information, or product.
Users taking actions become infected, and may influence others since 
such actions are publicly visible or pushed to the related users on purpose, by
the online service.
Thus we define a cascade $C$ for actions on a target 
as a temporal sequence 
\begin{equation}
    \nonumber
C=\{(v_1, t_1), (v_2, t_2),\cdots, (v_N, t_N) | t_1 \leq t_2 \leq \cdots \leq t_N\},
\end{equation}
where $v_i$ is the user who take the action at time $t_i$, 
and $N$ is the total number of infected users, $i.e.$ cascade size.
Since social networks are not always available or existing in many applications,
such as blogs, Yelp, Youtube, and online shopping, to make our model generally
applicable, network structures are ignored. That is to say, 
the influence between any pair of users is modeled, and a very small value of
influences can capture the underlying disconnections of the user network, and vice versa.
Moreover, in the following we can see that our model can honor 
social network as well in the objective. 
In addition, a special time $t_E>t_N$ is defined as 
the biggest time window which we observe cascade $C$ in,
namely the time when we take the snapshot for cascade $C$.

\subsection{Motivations}

Interpersonal influences are quite different especially 
when existing some social
relationships.
Most existing works intuitively model the interpersonal influences in a pair-wise 
manner with $n^2$ independent variables to learn, 
assuming that \emph{interpersonal influence between different pairs of users are independent
from each other, even if 
the influences are related to a common user}.
Such an overfitting problem becomes severe, when there
is not observed any propagation between a pair of connected users.
Taking Figure~\ref{fig:motivation} as an example,
two cascades \{$(a,t_a^1)$, $(c, t_c^1)$, $(e, t_e^1)$, $(f, t_f^1)$\} and 
\{$(a, t_a^2)$, $(b, t_b^2)$ $(e, t_e^2)$, $(f, t_f^2)$ $(d, t_d^2)$\} are observed. 
It is not seen that 
user $c$ is infected before user $d$ did, even though 
there is a social link from user $c$ to user $d$. In such a case, 
most existing models took the propagation probability or transmission rate between them as zero, or
some empirically small value~\cite{goyal2011data}, implying that it would never or
seldom  see successful propagation between the two users in the future.
Nevertheless, with the witness of propagation from users $c$ to
$e$ in one cascade, and from users $e$ to $d$ in another, user $c$ probably
influences user $d$ like the triangle pattern in friendship relations.
Thus, with the distributed representations of influence and susceptibility
defined for every user,
interpersonal influences can be correlated
by the shared representations of the same user.
As shown in the example of Figure \ref{fig:motivation},
the influences between user pairs $(c,d)$ and $(c,e)$ are
coupled with a shared representation of user $c$'s influence.
And the interpersonal influence from user $c$ to $d$
can be intuitively estimated by the learned representations of $c$'s influence and
$d$'s susceptibility,
other than a small empirical constant or zero.

At last, not all the users take actions in a cascade, so
let the total number of users be $M$, and 
there are always a large number of users immune to a 
contagion, $i.e.$, $M-N \gg N$, who are treated as negative cases 
and informative
to reflect the interpersonal influences from the infected users to them.
Without network constants, considering all the uninfected users
takes much more computational costs, even unable to tackle. 
Moreover, the severe imbalanced positive (infected) cases and negative
(uninfected) cases make the negative likelihood dominate the 
optimization of the whole objective, losing focus on
positive cases as the following.
\begin{equation}
    \nonumber
    \displaystyle\max \sum_c\ln\mathcal L = 
    \displaystyle\sum_c\sum^{N^c}\ln\mathcal L^c_{pos} +\sum_c\sum^{M^c-N^c}\ln\mathcal L^c_{neg} \text{(dominate)}
\end{equation}
where superscript $c$ is used to show that the values are related 
to cascade $C$. It is seen that the right term
in summation easily dominates the objective, since
$M$ is relatively very large.
So we use sampled users as negative cases. Nevertheless,
the infected frequency of a user indicates how 
easily she could get infected again.
Thus observing a frequently infected user immune
to a contagion, provides more information in 
the likelihood. And sampling negative cases
from the distribution of users' infected frequencies
is then a better choice for learning influences. 

\subsection{Survival Analysis Model}
We begin to briefly introduce the preliminary knowledge on 
Survival Analysis Model~\cite{Lawless2011,gomez2013modeling}.
We consider the happening time $T$ of a user taking the action 
as a continuous random variable, defined over $[0, \infty)$.
Let $f(t)$ and $F(t)$ denote the probability density function (p.d.f)
and the cumulative density function (c.d.f.) separately. And the probability
$Pr(T \leq t) = F(t)$.
So, the probability of a user not taking the action until time $t$ is
defined by the survivor function 
\begin{equation}
    S(t) = Pr(T \geq t) = 1 - F(t) = \int_t^\infty f(x) dx .\nonumber
\end{equation}
A hazard function $h(t)$ is defined as the instantaneously 
infecting rate in time interval $[t, t+\varepsilon)$,
where $\varepsilon$ is an infinitesimal elapsed time, given
a user survives until time $t$.
\begin{equation}
    \begin{split}
	\nonumber
    h(t) &= \lim_{\varepsilon \rightarrow 0} \frac{Pr(t \leq T < t+ \varepsilon |
    T \geq t)}{\varepsilon} = \frac{f(t)}{S(t)}.
    \end{split}
\end{equation}
Noticing that $f(t)=-S'(t)$ and $S(0)=1$ , the survivor function can be 
expressed as
\begin{equation}
    \nonumber
    \ln S(x) = -\int_0^t h(x) dx.
\end{equation}

\subsection{Modeling sentimental cascades}

With the analysis above, we model the interpersonal influence by two non-negative
$K \times D$ matrices $\mathbf I_i$ and $\mathbf S_i$ defined on each user $v_i$,
where $K$ is the number of sentiment classes, and $D$ is the 
dimension of users' representations on each sentiment class.
For a message with sentimental opinion, we define a one-hot vector
$o$ with $K$ dimensions, representing its exclusive sentiment class. 
Thus, for a cascade with sentiment $o$, 
the transmission rate function $\phi(\cdot)$ from users $v_j$ to $v_i$,
is defined as equation (\ref{eqtransrate}), which indicates the likelihood of
successful propagation between them. Although the original concept of 
transmission rate is not necessarily between 0 and 1, we
scale it for regularization.

\begin{equation}
    \label{eqtransrate}
    \phi(\mathbf I_j, \mathbf S_i, o ) = 1 - \exp\{- o^T\mathbf I_j \mathbf S_i^To\}
\end{equation}
where matrices $\mathbf I_j$ and $\mathbf S_i$ are parameters 
to separately capture the influence 
of user $v_j$ and the susceptibility of user $v_i$.
Let $\mathcal{H}_{ji}$ denote 
the set of parameters $\{\mathbf I_j, \mathbf S_i, o \}$ for simplification.
With transmission rate $\phi(\mathcal H_{ji})$, we can define the hazard function
from users $v_j$ to $v_i$ in Survival Analysis Model, at time $t$ as follows.
\begin{equation}
    \label{eqhazard}
    h(t | t_j; \phi(\mathcal H_{ji}) ) = \phi(\mathbf I_j, \mathbf S_i, o )\frac{1}{t-t_j+1},
\end{equation}
where $t-t_j+1$ depicts the hazard function monotonously decaying with the time elapsed
from $t_j$, and adding 1 avoids unbounded hazard rate 
due to a zero or infinitesimal value of $t-t_j$.
Noticing that equation (\ref{eqhazard}) holds only when $t \geq t_j$, we
define hazard rate $h(t | t_j; \phi(\mathcal H_{ji}) ) = 0$,
when $t < t_j$, namely, user $v_j$ has not been infected at 
time $t$.
Moreover, we can consider social network by defining hazard function
$h(t | t_j; \phi(\mathcal H_{ji}) ) = 0$ as well, if user $v_i$ and user
$v_j$ are not connected.

And then the survivor function  
$S(t|t_j; \phi(\mathcal H_{ji}) )$ of user $v_i$ surviving later than time $t$
 and under the influence of user $v_j$, satisfies
\begin{equation}
    \begin{split}
    \ln S(t|t_j; \phi(\mathcal H_{ji})) &= - \int_0^t h(x | t_j; \phi(\mathcal H_{ji})) dx \\
    &= \phi(\mathbf I_j, \mathbf S_i, o) \cdot \ln(t-t_j+1)
    \end{split}
\end{equation}
Finally, the probability density function of 
user $v_i$ happening (getting infected) at time $t$, given
user $v_j$ happening (infected) at time $t_j$ is calculated as
follows.
\begin{equation}
    \nonumber
    f(t | t_j; \phi(\mathcal H_{ji})) = h(t | t_j; \phi(\mathcal H_{ji})) S(t|t_j; \phi(\mathcal H_{ji})). 
\end{equation}

\begin{figure}[tb]
\centering
\includegraphics[width=0.37\textwidth]{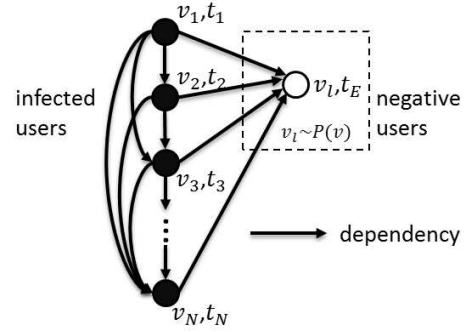}
\caption{A concise representation of dependencies of users' behaviors in the
model. Solid nodes are infected users, and hollow nodes are the immune ones (negative) 
to the contagion. Model parameters are omitted for simplicity.}
\label{figdependency}
\end{figure}

With the assumption that a user is only infected by one of the
previously infected ones \cite{gomez2013modeling}, the likelihood
of user $v_i, i>1$, being infected at time $t_i$ in a cascade is
\begin{eqnarray}
    \nonumber
    f(t_i | \bm t; \phi(\bm{\mathcal H})) =& \displaystyle\sum_{j: t_j<t_i} f(t_i | t_j; \phi(\mathcal H_{ji}))
    \prod_{k \neq j, t_k < t_i}S(t_i | t_k; \phi(\mathcal H_{ki})) \\
    \label{eqlikelihoodti}
    =& \displaystyle\sum_{j:t_j<t_i} h(t_i | t_j; \phi(\mathcal H_{ji}) \cdot 
    \prod_{k: t_k<t_i} S(t_i|t_k;\phi(\mathcal H_{ki})).
\end{eqnarray}
So the joint likelihood of observing the whole cascade, given 
the user $v_1$ firstly taking the action at time $t_1$ is 
\begin{eqnarray}
    \nonumber
    f(\bm t \setminus t_1 | t_1; \phi(\bm{\mathcal H})) =& \displaystyle\prod_{i>1}\sum_{j:t_j<t_i} 
    h(t_i|t_j;\phi(\mathcal H_{ji})) \cdot \\\nonumber
    &\displaystyle\prod_{k:t_k<t_i} S(t_i|t_k;\phi(\mathcal H_{ki})). 
\end{eqnarray}
Considering the negative cases that users are not infected at the end,
the probability of user $v_l$ surviving later than time $t_E$ is
\begin{equation}
    \nonumber
    S(t_E|\bm t; \phi(\bm{\mathcal H})) = \prod_{i: t_i\leq t_N} S(t_E | t_i; \phi(\mathcal H_{il})).
\end{equation}
And the log-likelihood of a cascade 
is as follows, considering negative cases.
\begin{eqnarray}
    \nonumber
    \ln \mathcal L(\mathbf{I,S}; o) =& \displaystyle \sum_{i>1} \ln \left ( \sum_{j:t_j<t_i} 
    \phi(\mathbf I_j, \mathbf S_i, o)\frac{1}{t_i - t_j + 1} \right ) - \\
    \nonumber
    &\displaystyle\sum_{i>1} \sum_{k:t_k<t_i} \phi(\mathbf I_k, \mathbf S_i, o)\cdot \ln(t_i-t_k+1) - \\
    \nonumber
    &\displaystyle\sum^L \mathbb E_{v_l \sim P(u)} \left[
    \sum_{j=1}^{N}\phi(\mathbf I_j, \mathbf S_l, o)\cdot \ln(t_E-t_j+1) \right]
\end{eqnarray}
There are a large number of negative users comparing to the number
of infected ones in a cascade. Maximizing the likelihood of all
the negative cases limits the scalability of our model, and
the imbalance between positive and negative cases may mislead the
optimization direction. Thus we sample $L$ users as negative cases according
to the distribution $P(u) \propto R_u^{3/4}$ \cite{mikolov2013distributed}, 
where $R_u$ is the frequency of user $u$ infected in cascades. 
It is worth noticing that sampling negative
cases are repeated in every optimization iteration to honor the 
expectation. To give a direct understanding of the likelihood, 
the dependencies are concisely represented in Figure \ref{figdependency}. 

Finally, the optimization problem of learning users' sentimental influences and susceptibilities 
\begin{subequations} 
    \label{eqopt}
\begin{eqnarray}
    \label{eqopt:a}
    \displaystyle \min_{\mathbf I, \mathbf S} &  \displaystyle -\sum_{C} \ln \mathcal{L}^c(\mathbf{I,S}; \bm o^c) 
    \\
    \label{eqopt:b}
    \displaystyle \textit{s.t. } & \bm I_{ki} \geq \mathbf{0}, \bm S_{ki} \geq \mathbf{0}, \forall k,i.
\end{eqnarray}
\end{subequations}
where superscript $c$ is used to indicate 
that the value or function are related to cascade $C$.

\subsection{Optimization}

Optimization algorithm is the key to learn the distributed
representations of users' influences. 
First of all, the gradients of transmission rate function
(\ref{eqtransrate}) on $\mathbf I_v$ and
$\mathbf S_u$ are $K\times D$ matrices.
\begin{eqnarray}
    \nonumber
    \frac{\partial\phi(\mathbf I_v, \mathbf S_u, o )}{\partial \mathbf{I}_v} &=&
      (1-\phi(\mathbf I_v, \mathbf S_u, o ))oo^TS_u \\\nonumber
    \frac{\partial\phi(\mathbf I_v, \mathbf S_u, o )}{\partial \mathbf{S}_u} &=&
      (1-\phi(\mathbf I_v, \mathbf S_u, o ))oo^TI_v
\end{eqnarray}
where only the $k$-th row in both matrices can have non-zero gradients, 
when a cascade belongs to the $k$-th sentiment class, $i.e$, $o_k=1$.
Furthermore, if user $v$ get infected in a cascade, $t_1\leq t_v\leq t_N$, 
the gradients of the log-likelihood 
on matrix $\mathbf I_v$ may have non-zero gradients.
And the gradients of the log-likelihood 
on matrix $\mathbf S_v$ may be non-zeros, 
if user $v$ is infected and $t_1<t_v\leq t_N$, or she is 
a negative user.
Otherwise, the gradients are always zeros.
As the negative cases for a cascade is repeatedly
sampled in every iteration, we define $[\mathbb V_s^c]_\tau$
as the set of negative users at the $\tau$-th iteration
of algorithm for cascade $C$.
\begin{equation}
    \nonumber
    [\mathbb V_s^c]_\tau = \{v_l \sim P(u)\}_L,
\end{equation}
where $L$ is the set size. 

Therefore, the gradients of the objective 
function (\ref{eqopt:a}) on matrices $\mathbf I_{v}$ and
$\mathbf S_{v}$ are as follows.
\begin{eqnarray}
    \nonumber
g_{I_v} &=& -\displaystyle\sum_c \mathbf 1(t_v^c \leq t_N^c) 
  \frac{\partial \mathcal L^c(\mathbf{I,S}; \bm o^c)}{\partial \mathbf I_v} \\\nonumber
g_{S_v} &=& -\displaystyle\sum_c \mathbf 1(t_1^c < t_v^c \leq t_N^c)  
  \frac{\partial \mathcal L^c(\mathbf{I,S}; \bm o^c)}{\partial \mathbf S_v} +
     \sum_c \mathbf 1(v \in [\mathbb V_s^c]_\tau) \cdot \\\nonumber
     & &  \displaystyle\sum_{j=1}^{N^c}(1-\phi(\mathbf I_j, \mathbf S_v, o^c))
          \cdot \ln(t_E^c-t_j^c+1)o^c{o^c}^T\mathbf I_j
\end{eqnarray}
where 
$\mathbf 1(\cdot)$ is an indicator function, 
outputting 1 if the argument is true, and 0 otherwise.
$g_{I_v}$ and $g_{S_v}$ are
$K \times D$ matrices, containing partial derivations of
objective function (\ref{eqopt:a}) on each elements 
of matrices $\mathbf I_v$ and $\mathbf S_v$ separately.

The framework of Stochastic Gradient Decent (SGD) 
over shuffled mini-batches is employed for 
efficient optimization.
The mini-batch size is set 12 cascades.
In order to solve the non-negative constraints on parameters,
Projected Gradient (PG) \cite{lin2007projectgradient} is 
used to adjust the gradients.
Let the parameter updates be $\Delta \mathbf I_v$ and 
$\Delta\mathbf S_v$ for each user $v$.
And matrices $\mathbf {I,S} \in \mathbb R_{+}^{K\times DM}$ are the 
concat of $\mathbf I_v$ and $\mathbf S_v$ for all users $v$.
$M$ is the user count as defined previously.
Thus the updates will be reduced by a rate $0<\beta<1$, namely,
$\beta\Delta\mathbf I_v$ and $\beta\Delta\mathbf S_v$, if
the following condition does not hold.
\begin{equation}
    \label{eqcondPGD}
    \mathcal O([\mathbf E]_{\tau+1}) - \mathcal O([\mathbf E]_\tau) \leq
       \sigma\cdot Tr(\nabla \mathcal O([\mathbf E]_\tau)^T([\mathbf E]_{\tau+1}-[\mathbf E]_\tau))
\end{equation}
where $[\cdot]_\tau$ means the parameter in the $\tau$-th iteration.
With $\mathbf E = \{\mathbf I, \mathbf S\} \in \mathbb R^{K \times 2DM}$,
$\mathcal O(\mathbf E)$ is the simplified representation of objective function
(\ref{eqopt:a}).
$Tr(\cdot)$ is the trace of a matrix, and $\sigma$ is a constant
between 0 and 1. 

Moreover, since deciding learning rate is not trivial, so
we choose Adadelta \cite{zeiler2012adadelta} to 
adaptively tune the learning rate.
Let $\rho$ be decay rate and $\epsilon$ be a small constant.
The accumulate gradients are
\begin{eqnarray}
    \nonumber
    E[g_{I_v}^2]_\tau &=& \rho E[g_{I_v}^2]_{\tau-1} 
       + (1-\rho)[g_{I_v}^2]_\tau \\\nonumber
    E[g_{S_v}^2]_\tau &=& \rho E[g_{S_v}^2]_{\tau-1} 
       + (1-\rho)[g_{S_v}^2]_\tau \\\nonumber
    E[{\Delta\mathbf I_v}^2]_\tau &=& \rho 
       E[{\Delta\mathbf I_v}^2]_{\tau-1} 
       + (1-\rho)[{\Delta\mathbf I_v}^2]_\tau\\\nonumber
    E[{\Delta\mathbf S_v}^2]_\tau &=& \rho 
       E[{\Delta\mathbf S_v}^2]_{\tau-1} 
       + (1-\rho)[{\Delta\mathbf S_v}^2]_\tau
\end{eqnarray}
And with the definition of 
function $RMS[x]_\tau=\sqrt{E[x^2]_\tau+\epsilon}$,
the update values are calculated as 
\begin{eqnarray}
    \nonumber
    [\Delta\mathbf I_v]_\tau &=& -\displaystyle\frac{RMS[\Delta\mathbf I_v]_{\tau-1}}
    {RMS[g_{I_v}]_\tau} [g_{I_v}]_\tau \\\nonumber
    [\Delta\mathbf S_v]_\tau &=& -\displaystyle\frac{RMS[\Delta\mathbf S_v]_{\tau-1}}
    {RMS[g_{S_v}]_\tau} [g_{S_v}]_\tau 
\end{eqnarray}

Let the project function $\psi(x)$ be defined as projecting
$x$ into non-negative space, namely, $\psi(x)=0$ if $x<0$;
otherwise $\psi(x)=x$.
Therefore, the algorithm of learning users' sentimental
influences is listed in Algorithm \ref{alglearning}.

\begin{algorithm}
\caption{Algorithm of learning users' sentimental influences.}
\label{alglearning}
\begin{algorithmic}
\STATE {\bf Given} $0<\rho, \beta<1$, constants $\sigma$ and $\epsilon$;\\
\qquad initialized parameters $\mathbf I_v$ and $\mathbf S_v$ for each user $v$;\\
\qquad Cascade set $\mathbb C$.
\STATE
\STATE Iteration index $\tau := 0$;\\
\STATE $E[g_{I_v}^2]_0, E[g_{S_v}^2]_0, E[{\Delta\mathbf I_v}^2]_0, E[{\Delta\mathbf S_v}^2]_0 = 0$;
\REPEAT
   \STATE Randomly shuffle $\mathbb C$; 
   \STATE Split $\mathbb C$ into groups by mini-batch size;
   \FOR {each group}
        \STATE Compute gradients $[g_{I_v}]_\tau, [g_{S_v}]_\tau$;
	\STATE Accumulate gradients and updates:\\
	\quad $E[g_{I_v}^2]_\tau, E[g_{S_v}^2]_\tau, 
	E[{\Delta\mathbf I_v}^2]_\tau, 
	E[{\Delta\mathbf S_v}^2]_\tau$;
	\STATE Parameter update values:\\
        \quad $[\Delta\mathbf I_v]_\tau,
	[\Delta\mathbf S_v]_\tau$
	\STATE Update $[\mathbf I_v]_{\tau+1} = \psi([\mathbf I_v]_{\tau}+[\Delta\mathbf I_v]_\tau)$;\\
	\STATE Update $[\mathbf S_v]_{\tau+1} = \psi([\mathbf S_v]_{\tau}+[\Delta\mathbf S_v]_\tau)$;\\
	\WHILE{\NOT Condition (\ref{eqcondPGD})}
	   \STATE decreasing update values:\\
	   \quad $[\Delta\mathbf I_v]_\tau = \beta [\Delta\mathbf I_v]_\tau$, 
	         $[\Delta\mathbf S_v]_\tau = \beta [\Delta\mathbf S_v]_\tau$
	   \STATE Update $[\mathbf I_v]_{\tau+1} = \psi([\mathbf I_v]_{\tau}+[\Delta\mathbf I_v]_\tau)$;\\
	   \STATE Update $[\mathbf S_v]_{\tau+1} = \psi([\mathbf S_v]_{\tau}+[\Delta\mathbf S_v]_\tau)$;\\
	\ENDWHILE
        \STATE $\tau := \tau+1$
   \ENDFOR
\UNTIL parameters converged, or maximum epoch. 
\end{algorithmic}
\end{algorithm}

\section{Evaluations}
\label{secexperiments}
Microblog data is used to evaluate our model.
To make the application more general, we
assume that the retweeting relations and following
relations are not available in the evaluations, 
only keeping the temporal sequence of users taking 
actions, $i.e.$, retweet, and their infected 
times as the dataset.
We then demonstrate the performance of our model at
the well-known tasks, by comparing to the state-of-the-art models, 
and the learned sentimental influences are
analyzed as well.

\subsection{Data Description}

Several strategies are taken to collect Microblog data from Sina Weibo
\footnote{Sina Weibo (http://www.weibo.com) is the biggest site for Microblog 
service in China.}.
We initially collected about 315.6 million records 
including posting, retweeting, and mentioning 
messages between Nov 1st, 2013 
to Feb 28, 2014 from the timeline of 
312,000 users sampled from Sina Weibo database.
Since emoticons in cascade messages are usually used as the sentiment indicator, 
we filtered the messages with frequently used emoticons and active users, 
and crawled the full records of retweeting cascades
of those remaining messages.
Emoticons are split into positive sentiment set and negative sentiment set according
to a dictionary of emoticons. And sentiments of messages are intuitively assigned
according to emoticons in our experimental settings. Otherwise, one can use
any reliable sentiment classifier, such as OpinionFinder~\cite{Choi2005} to decide the sentiments.
Meanwhile, retweeting relations are also extracted from the auto-generated contents, which
help to preprocess data, and are used for ground truth in later evaluations.
And we define the activeness $A_v$ of a user $v$ as the summation of 
the frequency of user $v$ getting infected (retweeting others), $A_{\cdot v}$, 
and that of user $v$ influencing others (being retweeted), $A_{v \cdot}$, in current cascades.
\begin{equation}
    \nonumber
    A_v = A_{\cdot v} + A_{v \cdot}.
\end{equation}
Afterward, in a way of ``onion peeling'', we repeated to delete for each cascade, the 
records of the users with activeness less than 5, and so did those of the 
users retweeting them. In each iteration, the cascades of sizes less than 8 are
deleted as well, since very short cascades are considered as accidents. 

\begin{table}[htbp]
  \centering
  \caption{Data Statistics}
  \vspace{2pt}
  \subtable[]{ 
    \begin{tabular}{lllll}
    \Xhline{1.2pt}
              & Total & Total      & \multicolumn{2}{c}{Sentiment} \\
    \cline{4-5}
    Time span & users & cascade size &Positive & Negative  \\
    \hline
    10/31/13  &       &        &        &  \\
    -3/3/14   & 6219  & 44021  &  325   & 412   \\
    \Xhline{1.2pt}
    \end{tabular}%
    }
    \subtable[]{ 
    \begin{tabular}{lllll}
    \Xhline{1.2pt}
    \multicolumn{2}{c}{User activeness} & & \multicolumn{2}{c}{Cascade size} \\
    \cline{1-2} \cline{4-5}
    median & mode & & median & mode \\
    \hline
    5     & 4     && 37      &10  \\
    \Xhline{1.2pt}
    \end{tabular}%
    }
  \label{tbdata}%
\end{table}%

\begin{figure}[t]
\centering
\includegraphics[width=0.42\textwidth]{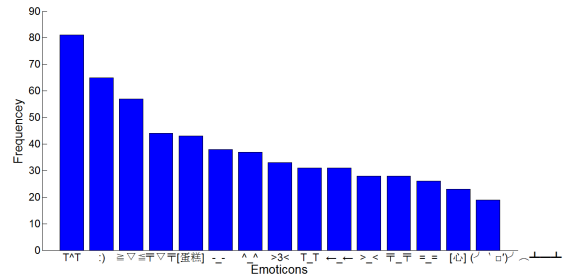}
\caption{The distribution of emoticon frequency.}
\label{figemoticons}
\end{figure}

With such a heuristic way, we finally get a set of cascades
over a virtual community of active users from Oct 31, 2013 to Mar 3, 2014.
As listed in Table \ref{tbdata}(a), there are 6,219 users, and 44,021 cascade
records totally. The number of cascade messages with positive emoticons is
325, and the number of those with negative ones is 412, keeping a balanced observations
for learning sentimental influences.
And Figure \ref{figemoticons} illustrates the distributions
of top frequently used emoticons in the messages of cascades, 
indicating their positive sentiments or negative sentiments.  
Furthermore, the median and mode values of the distribution of 
users' activeness are 5 and 4 separately as in Table \ref{tbdata}(b), indicating
that users' behaviors are not rarely observed in our dataset to guarantee a successful
learning. And it also gives the median and mode of the distribution of cascade sizes
as well, showing the sufficiency of involved users in a cascade.
The cascades in the dataset are evenly split into 
10 groups, and 10-fold cross testing are used for evaluations, alternatively with 9 of 10 groups as
training, and the remaining one as testing.

\subsection{Evaluation Models}
In the experiments, we choose the following models for comparison.
\begin{itemize}
    \item[-]\textbf{CT Bernoulli} and \textbf{CT Jaccard} models~\cite{goyal2010learning}: 
	They are continuous time models that 
	the propagation probability $P_{ab}$ from user $a$ (infected) to $b$ 
	decays with the elapsed time. For a fair comparison, we use 
	the same decaying function, $i.e.$, $P_{ab}=P_{ab}^0/{(t_b-t_a+1)}$, and
	the same assumption that a user is only infected by one of the infective users.
	CT Bernoulli model assumes that an initial propagation probability $P_{\cdot}^0$
	follows Bernoulli distribution, i.e. the fraction of number of successful 
	propagation over the total number of trials, from one user to another. 
	And CT Jaccard model defines an initial propagation probability  $P_{\cdot}^0$
	in a form of Jaccard Index, which is the 
	number of successful propagation divided by the total 
	number of cascades with at least one infected between a pair of users.
	Since there only observes a temporal sequence of users getting infected in
	training dataset, we assume that successful propagation takes place 
	from every earlier infected users to the current one.
    \item[-] \textbf{NetRate}\cite{GomezRodriguez2011}: It directly 
	define a scalar parameter as interpersonal influence between a
	pair of users, and learned them with Survival model.
	Since Jaccard Index was reported as a better estimator of
	propagation probability \cite{goyal2010learning}, 
	we use Jaccard Index to initialize the transmission rates
	at the beginning of the learning stage, to get a better fine tune. 
    \item[-] \textbf{CT LIS}: We ignore the differences of latent influence and susceptibility
	on sentiments of cascade messages, and define two $D$-dimensional 
	vectors, $I_v$ and $S_v$, for transmission rate function $\phi(\cdot)$ instead.
	Such parameters were ever defined by \cite{wang2015learning}, which used a static way to
	model the orders of users' behaviors. So we use ``CT LIS'' to indicate our upgrade version
	for continuous time model.
    \item[-] \textbf{Sent LIS}: It is our model that learns sentimental influences considering
	all the negative cases. And we use ``Sent LIS (neg sample)'' to indicate ours with
	negative sampling.
\end{itemize}

\begin{table*}[tb]
  \vspace{4pt}
  \centering
  \caption{Average MRRs and AUCs of PCD task for 10-fold cross testing.}
    \begin{tabular}{llllllll}
    \Xhline{1.2pt}
    \multicolumn{2}{c}{} & CT Bernoulli & CT Jaccard & NetRate (Jaccard) & CT LIS & Sent LIS  & Sent LIS (neg sample) \\
    \hline
    \multirow{2}[4]{*}{MRR} & Average & 0.0062  & 0.0064  & 0.0071  & 0.0196  & \textbf{0.0216} & \textbf{0.0265}  \\
          & SD    & $\pm$0.0029  & $\pm$0.0036  &$\pm$ 0.0038  &$\pm$ 0.0039  &$\pm$ 0.0033  &$\pm$ 0.0044  \\
    \multirow{2}[4]{*}{AUC} & Average & 0.8732  &    0.8621  &    0.8718  &    0.8793  & {\bf 0.8992 } & {\bf 0.8983 }  \\
          & SD    & $\pm$ 0.0658  & $\pm$ 0.0802  & $\pm$ 0.0730  & $\pm$ 0.0207  & $\pm$ 0.0152  & $\pm$ 0.0156  \\
    \Xhline{1.2pt}
    \end{tabular}%
  \label{tbpcd}%
\end{table*}%

\begin{table*}[phtb]
  \centering
  \caption{Average accuracies and MRRs of WBR task for 10-fold cross testing.}
    \begin{tabular}{llllllll}
    \Xhline{1.2pt}
    \multicolumn{2}{c}{} & CT Bernoulli & CT Jaccard & NetRate (Jaccard) & CT LIS & Sent LIS  & Sent LIS (neg sample) \\
    \hline
    \multirow{2}[4]{*}{Acc} & Average & 0.1221  & 0.3000  & 0.3005  & \textbf{0.4123 } & 0.3840  & \textbf{0.3980 } \\
          & SD   &$\pm$ 0.0365  &$\pm$ 0.0964  &$\pm$ 0.0961  &$\pm$ 0.0874  &$\pm$ 0.1255  &$\pm$ 0.1392  \\
    \multirow{2}[4]{*}{MRR} & Average & 0.2592  & 0.4349  & 0.4354  & 0.4696  & \textbf{0.4822 } & \textbf{0.4920 } \\
          & SD   &$\pm$ 0.0703  &$\pm$ 0.1275  &$\pm$ 0.1273  &$\pm$ 0.0876  &$\pm$ 0.1269  &$\pm$ 0.1348  \\
    \Xhline{1.2pt}
    \end{tabular}%
  \label{tbWBR}%
\end{table*}%

\subsection{Tasks and evaluation metrics}
The following tasks are used 
to evaluate the effectiveness of our learned 
sentimental influences and the improvements 
comparing to the other models.
And the metrics for each task are introduced as well.

\textbf{PCD: predicting cascade dynamics.} 
The happening times and infected users of cascade dynamics 
are both predictable by our model. 
However, in order to make the task simple and easy to 
evaluate, we design the task
that aims at predicting whether a user $v$ will be infected
at a given time $t$, knowing the previous truth, $i.e.$ the users who have been infected,
and their happening times before time $t$.
Thus on one hand, the task can be treated as a set of binary classification
problems, and we evaluate the results, 
with the infected users as the positive cases,
and finally uninfected users until time $t_E$ as the negative ones.
As for the positive cases, the likelihood of an infected user $v$ at
given time $t_v$, is given by 
$f(t_v |\bm t; \phi(\mathcal{\bm H}_{\cdot v}))$.
The likelihood of a negative user $v_l$,
if she had been infected right after the positive ones, $i.e.$, at
time $t_N+\epsilon$, would be calculated as
    $f(t_N+\epsilon |\bm t; \phi(\mathcal{\bm H}_{\cdot l}))$,
where $\epsilon$ is a very small constant.
Thus with the likelihood values for all the users, 
true positive (TP) rate and false positive (FP) rate 
can be calculated given any threshold.
And then AUC (the area of under the ROC
curve) can be evaluated as ~\cite{Fawcett2006}, 
where ROC is drawn with TP rate and FP rate as the coordinates.

On the other hand, given time $t$, and the observation of cascades before
that time, we can calculate the infected likelihood for candidates $v$,
by $f(t |\bm t; \phi(\mathcal{\bm H}_{\cdot v}))$.
Thus with ranking the candidates with their likelihood values, 
the top ones are the most probably infected, 
and a well-performed model can give a high rank to those
users happened at the moment.
In such a way, Mean Reciprocal Rank (MRR)~\cite{voorhees1999trec} 
for rankings at all times of users getting infected in cascades
is calculated as the metric.

\textbf{WBR: who will be retweeted.}
Microblog users get infected and take actions to 
retweet the message from one of their followees 
who posts or retweets it previously.
Thus the task predicting ``who will be retweeted'' is a way to examine interpersonal
influence under quantitative understanding. 
In the scene of multi-exposures,
high interpersonal influence will have high probability to be
forwarded. 
As such, given $(v_i,t_i)$, namely, user $v_i$ happened at time $t_i$,
the infective user that $v_i$ retweets is 
\begin{equation}
    \nonumber
    arg\max_{j:t_j<t_i} f(t_i | t_j; \phi(\mathcal H_{ji})) 
\end{equation}
We therefore
deal with the prediction task as a ranking problem of interpersonal
influence. The user with higher rank is more probable to be 
retweeted. We evaluate the prediction performance by metrics of average
\textit{Accuracy} (Acc) of top-one prediction and MRR. 
The ground truth of retweets can be extracted from the content of Microblog messages.
Larger values of Acc and MRR indicate better
predictions. 

\textbf{CSP: Cascade size prediction.}
Cascade size prediction, as a key part of influence maximization and viral
marketing, is one of the most important applications based on modeling
cascade dynamics. 
In our settings of CSP task, we choose the first $P$ users and acting times 
of each cascade as the initialization, and predict the cascade size at time
$t_N$, $t_N > t_P$, where $t_P$ is the actinig time of the $P$-th user $v_P$.
The simulatioin method is used to predict the cascade size by dynaimics models.
The prediction time span $t_N-t_P$ is evenly splited and marked by time scales.
Thus starting after time $t_P$,  
an infected user $u$ tries to influence
an uninfected user $v$ at each time scale $\tau_{i} > t_P$,
with the probability 
\begin{equation}
    \nonumber
Pr(T \leq \tau_{i} | t_u; \phi(\mathcal H_{u,v})) =
\displaystyle \frac{\int_{\tau_{i-1}}^{\tau_{i}} f(t|t_u; \phi(\mathcal H_{u,v})) dt}
{S(\tau_{i-1} | t_u; \phi(\mathcal H_{u,v}))}.
\end{equation}
And if user $v$ is infected at time sacle $\tau_i$ with such a sampling, 
she will be added as infected users at the following time scales.
The simulatioins are repeated, and 
the average cascade size is reported as the prediction.
Thus the predction can be evaluated by \textit{mean absolute
percentage error} (MAPE), 
where a smaller value indicates a better prediction.

\subsection{Evaluation results.}

As the description of dataset, we split the whole datasets into 10 groups for
cross testing. Thus each experiments are repeated 10 times, and the average
metrics and the Standard Deviation (SD) are reported.
And the dimension of users' representations on a sentimental polarity is 
$D=8$ in the following evaluations for computational efficiency.

\begin{figure}[hbt]
\centering
\includegraphics[width=0.49\textwidth]{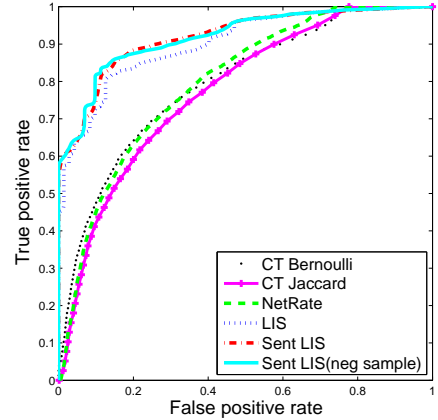}
\caption{The ROC curves of evaluation models on PCD task.}
\label{figroc}
\end{figure}

\textbf{PCD}: Figure \ref{figroc} illustrates the ROC curves of the 
evaluation models for one of the 10-fold cross tests.
It visually shows that
our models ``CT LIS'', ``Sent LIS'' and ``Sent LIS (neg sample)'' 
can achieve better performance in 
the formulation of binary classification. And NetRate with
Jaccard Index as parameter initialization improves the 
performance of ``CT Jaccard'' model.
As for the over all evaluations for 10-fold cross tests,
Table \ref{tbpcd} lists the average results and SDs of
all the models, with the best and the second best MRRs and
AUCs in bold text.
It is seen that our model achieves 0.0216 and 0.0265 in the 
metric of MRR, overwhelming other models 
with significance test, p-value < 0.01. And our negative
sampling model get the best, thanks to its effort in
balancing positive and negative cases.
By examining the results generated from ``CT Bernoulli''
and ``CT Jaccard'', it shows a consistent
result that Jaccard Index can beat the Bernoulli model
in the estimation of propagation probability, 
as \cite{goyal2010learning} reported.
In the measurement of binary classification, 
``Sent LIS'' and ``Sent LIS (neg sample)'' both outperforms
the others in AUC, which are 0.8992 and 0.8983 separately, 
with the former achieving a slightly better result.
Besides, the machine learning model NetRate can 
further tune the Jaccard Index to achieves better 
MRR and AUC values.
Most important of all, 
in both ranking and classification formulations 
of predicting cascade dynamics,
it is worth noticing that 
pair-wise models, namely, ``CT Bernoulli'', ``CT Jaccard''
and NetRate limits their performance, comparing to
the proposed models that learning distributed representations
of users, showing our advantages in the remission of overfitting
and model complexity reduction.

\begin{figure*}[htb]
\centering
   \subfigure[Our transmission rates on positive sentiment v.s. that of NetRate]
   {\includegraphics[width=0.24\textwidth]{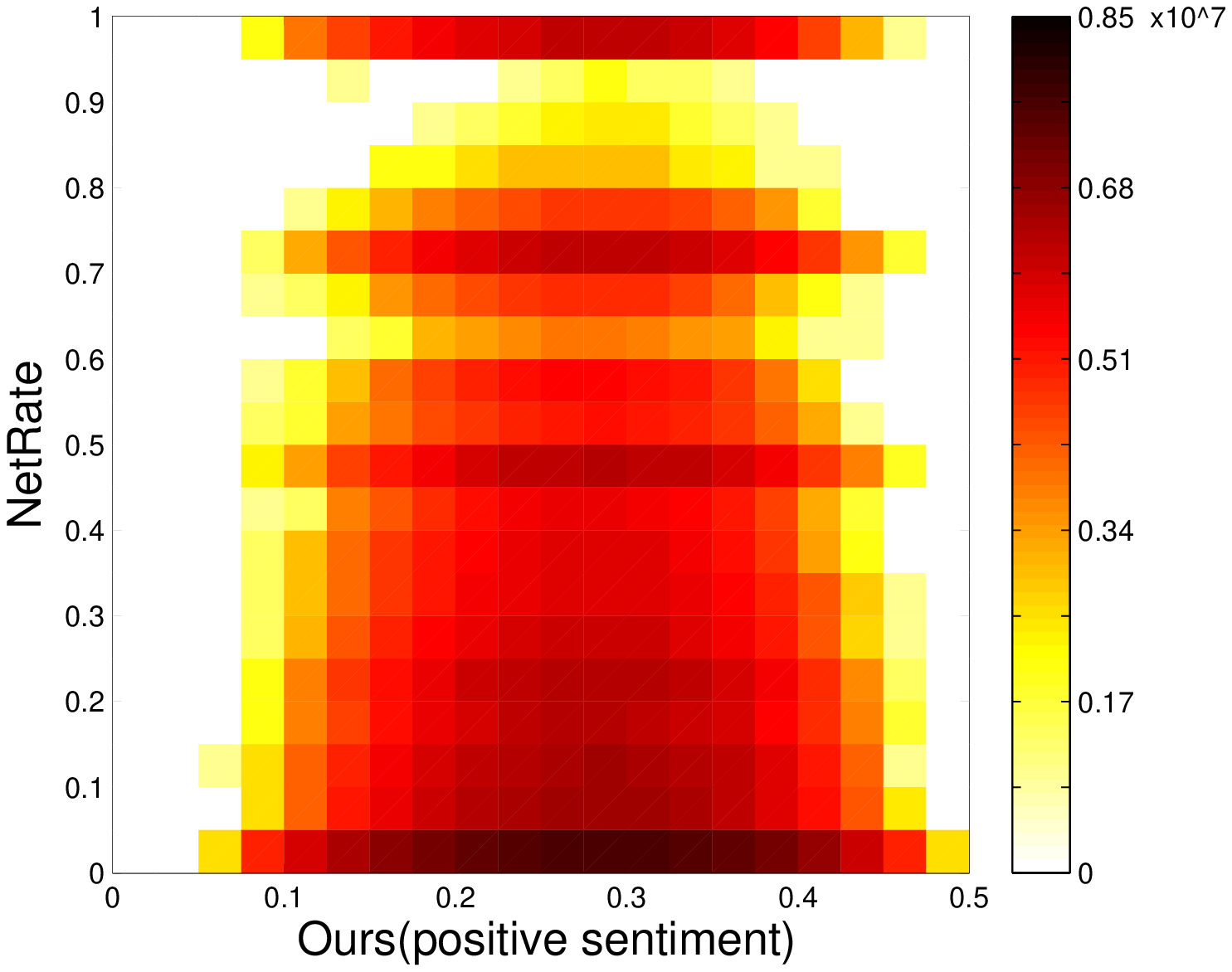}}
   \subfigure[Our transmission rates on negative sentiment v.s. that of NetRate]
   {\includegraphics[width=0.24\textwidth]{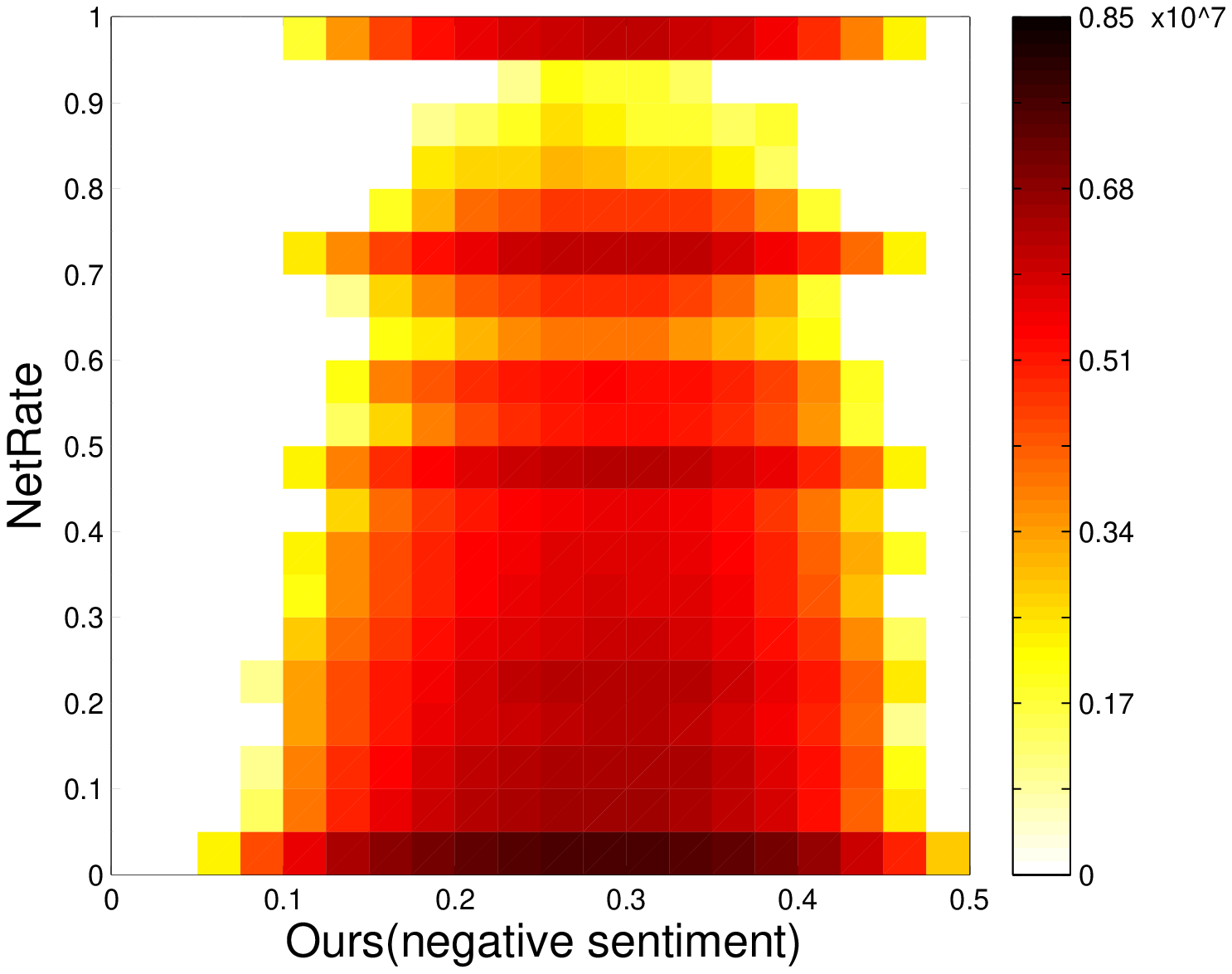}}
   \subfigure[Our transmission rates on positive sentiment v.s. that of Jaccard]
   {\includegraphics[width=0.24\textwidth]{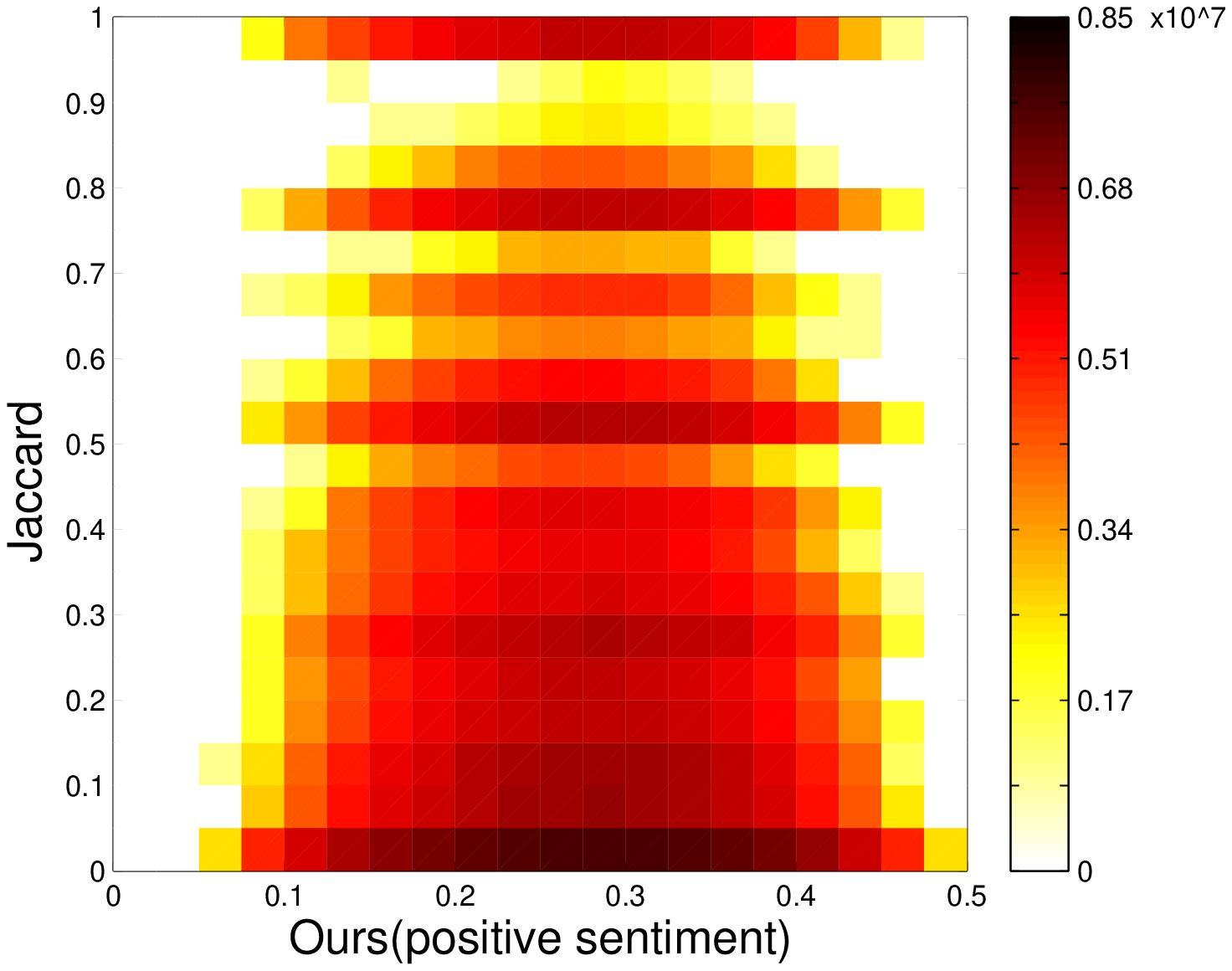}}
   \subfigure[Our transmission rates on negative sentiment v.s. that of Jaccard]
   {\includegraphics[width=0.24\textwidth]{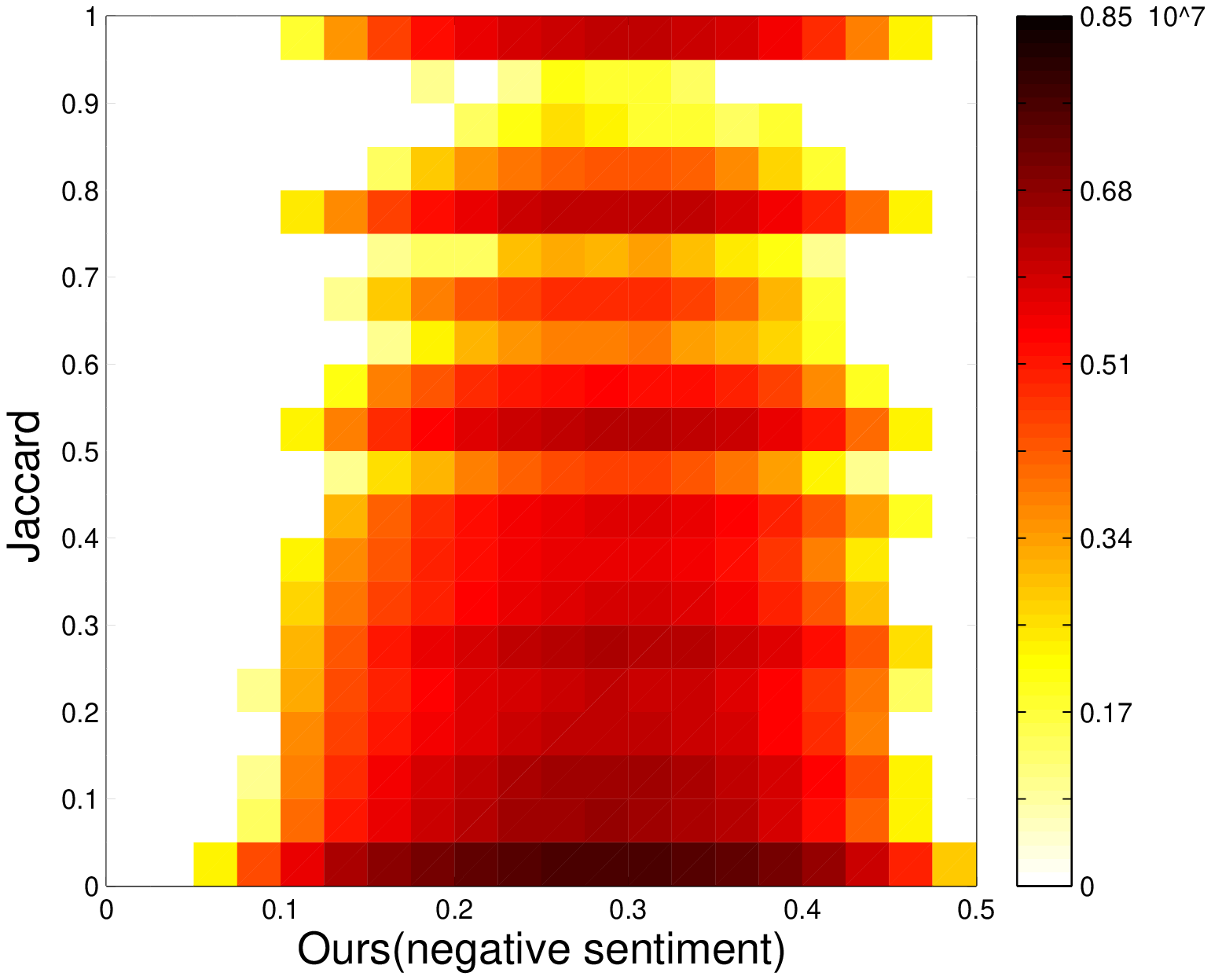}}
\caption{Analysis of transmission rates.}
\label{figtransrate}
\end{figure*}

\textbf{WBR}: With the extraction of retweeting relations from retweet content, 
the evaluation results of WBR task are reported in Table~\ref{tbWBR},
based on the ground truth.
The top-one accuracies (Acc) and MRRs of all cascades 
are averaged for the 10-fold cross
tests as well, and the significance is tested.
The bold numbers are the best and second best performances.
Again we can see that with distributed representations of users,
``CT LIS'', ``Sent LIS'', and 
``Sent LIS (neg sample)'' outperforms
the pair-wise models that suffer from the overfitting problems
on unobserved propagation pairs.
Compared with NetRate, the three LIS models improves
37.2\%, 27.8\% and 32.4\% separately 
on accuracies of predicting
``who will be retweeted'', while increasing the
MRRs by 7.9\%, 10.7\% and 13.0\% separately.
Besides, with the comparison among
pair-wise models, ``CT Jaccard'' still
takes her advantages to ``CT Bernoulli'' in
both metrics, and ``NetRate (Jaccard)'' is
the best of the three, thanks to the machine
learning with Survival model.
At last, with negative sampling, ``Sent LIS (neg sample)''
can on one hand balance the positive cases and negative
cases, and on the other consider the information
of negative cases in an expectation, resulting 
a better choice of decent gradient.
And in turn, it achieves a better performances 
in both accuracies and MRRs than ``Sent LIS'',
which testifies the advantages of our model.

\begin{table*}[phtb]
  \centering
  \caption{Average MAPE of CSP task for 10-fold cross testing.}
    \begin{tabular}{llllllll}
    \Xhline{1.2pt}
    \multicolumn{2}{c}{} & CT Bernoulli & CT Jaccard & NetRate (Jaccard) & CT LIS & Sent LIS  & Sent LIS (neg sample) \\
    \hline
    \multirow{2}[4]{*}{MAPE} & Average & 
    0.7199  & 0.7105  & 0.7109  & \textbf{0.6259}  & \textbf{0.6259} & \textbf {0.6362} \\
          & SD   &$\pm$ 0.0270  &$\pm$ 0.0333  &$\pm$ 0.0350  &$\pm$ 0.0883  &$\pm$ 0.1458  &$\pm$ 0.0.2252 \\
    \Xhline{1.2pt}
    \end{tabular}%
  \label{tbCSP}%
\end{table*}%

\textbf{CSP}: In the experiments, we choose first $P=10$ infected 
users as the initialization for prediction, and the times of
simulatioins for each cascade is 100 for efficiency.
Thus the averaged cascade sizes are reported with 10-cross
testing for all methods in Table \ref{tbCSP}.
It is seen that ``CT LIS'', ``Sent LIS'' without and with
negative sampling outperform other pair-wise models,
achieving 0.6259, 0.6259 and 0.6362 separately in MAPE. Moreover,
compared to the best-performed pair-wise model, we reduce
MAPE by more than 10.46\%,
which shows the advantage of our 
learned representations of users' influences in cascade size prediction.

Nevertheless, to show the differences of 
transmission rates learned by our model ``Sent LIS (neg sample)'' 
and pair-wise model NetRate, we separately calculate
the transmission rates of ours on positive sentiment and negative sentiment,
by latent sentimental influence and susceptibility matrices of equation
(\ref{eqtransrate}).
For each pair of users, there is a point with our transmission rate as X-coordinates,
and that of NetRate as Y-coordinates. And we count the number of points 
falling in each lattice cell, as illustrated in Figure \ref{figtransrate} (a) and (b),
which cells are colored from cold color to warm color 
based on the point counts.
Thus it is seen that a very warm and long line lying on the X-axis from
0.1 to 0.4 for both figures of positive sentiment and negative sentiment.
It tells that a lot of overfitting transmission rates by NetRate 
assigning a zero or small constant, can be estimated by the distributed
representations of users, which varies between different user pairs.
Besides, the higher transmission rates from NetRate can also have a
discriminative distribution in the transmission rates of ours, as
those horizontally aligned warm cells shows.
And the same solution can also be concluded from Figure \ref{figtransrate}
(c) and (d).
All above gives an evidence that our learned sentimental
influences have more abilities to discriminate in the influential 
and the susceptible, resulting good performances in the 
above evaluation tasks.

\subsection{Analysis of users' sentimental influences and susceptibilities} 

\begin{figure*}[tb]
\centering
   \subfigure[Positive influence v.s. positive susceptibility]
   {\includegraphics[width=0.29\textwidth]{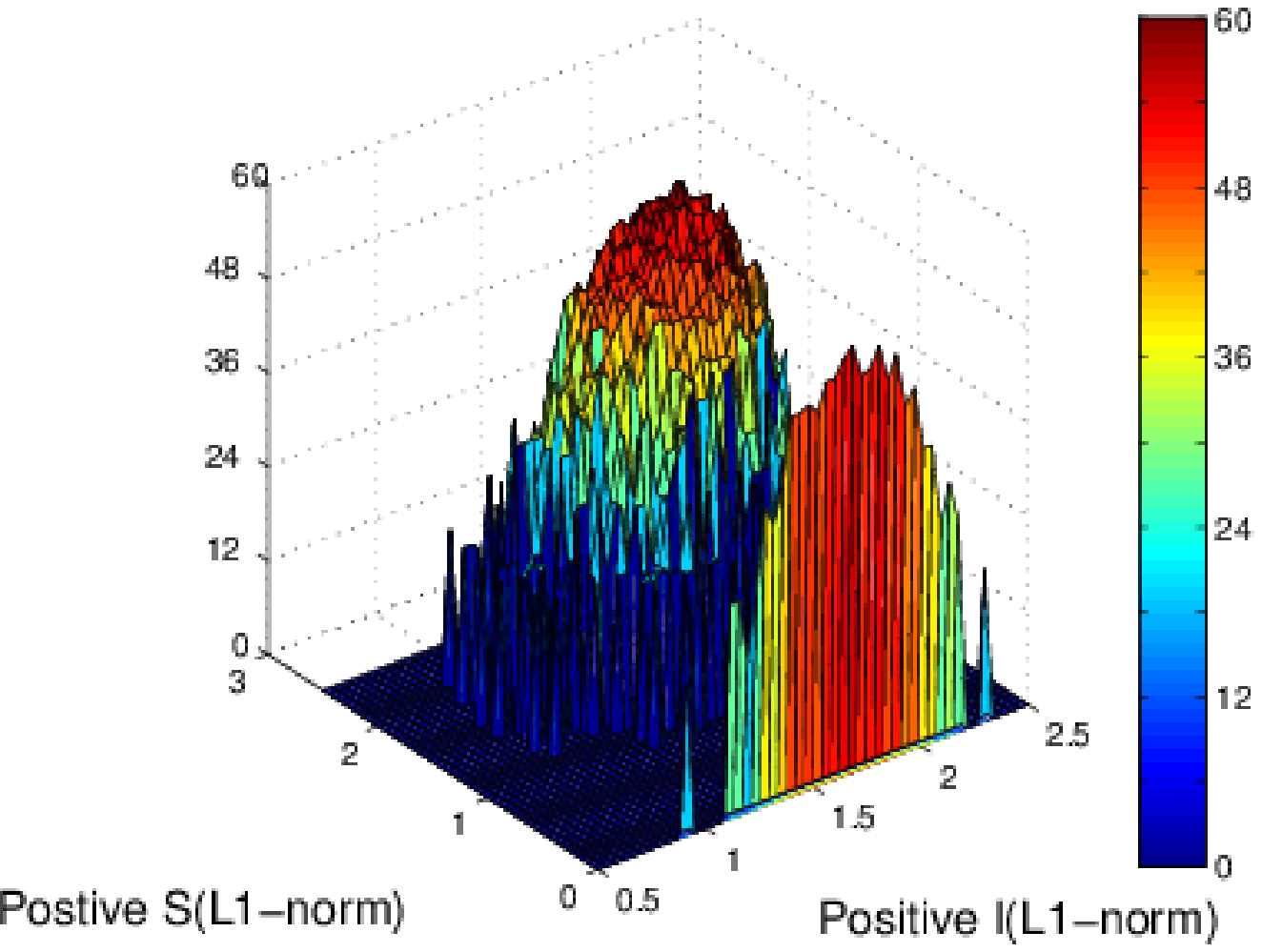}}\qquad
   \subfigure[Negative influence v.s. negative susceptibility]
   {\includegraphics[width=0.29\textwidth]{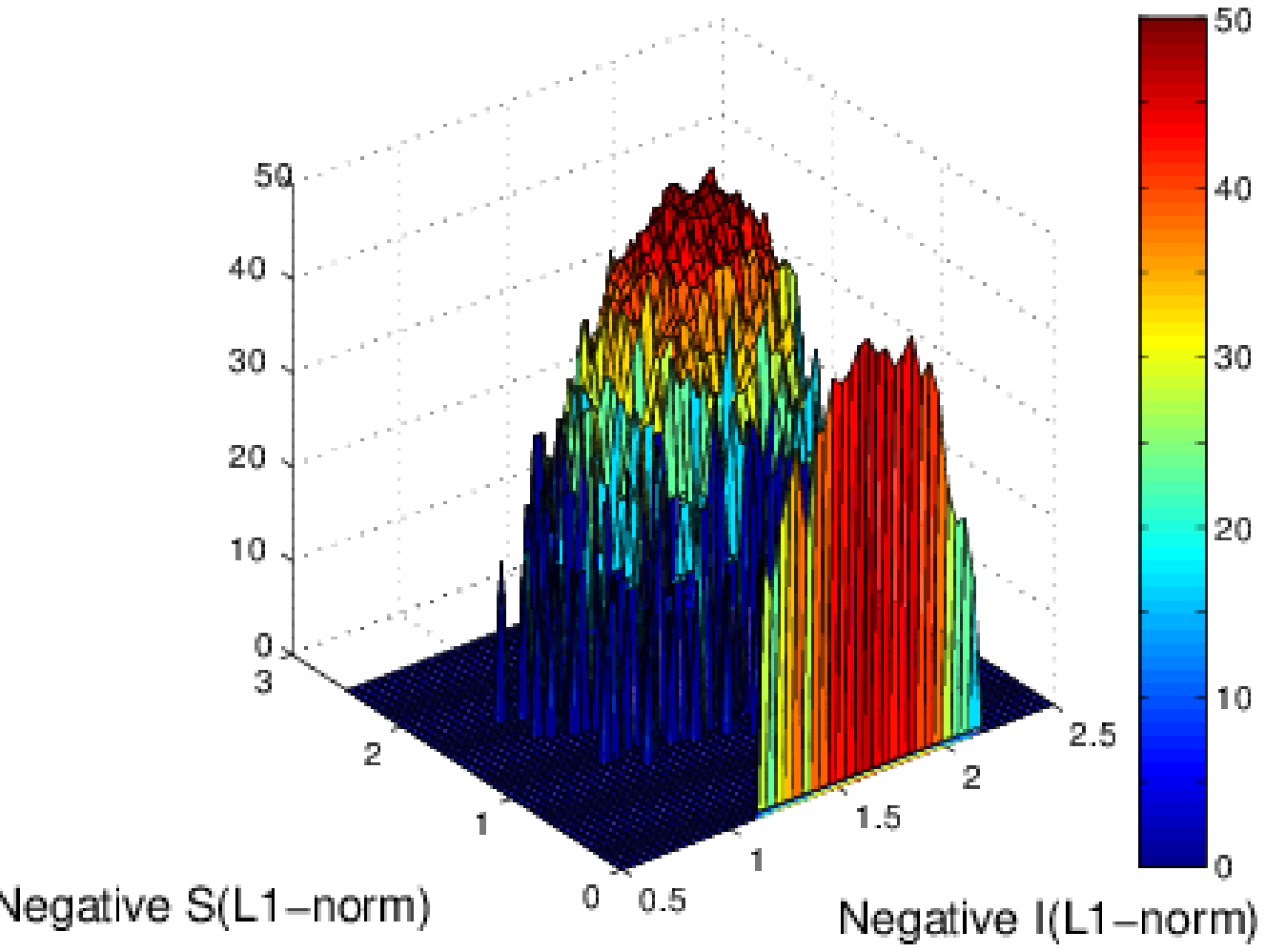}}\\
   \subfigure[Positive influence v.s. negative influence]
   {\includegraphics[width=0.25\textwidth]{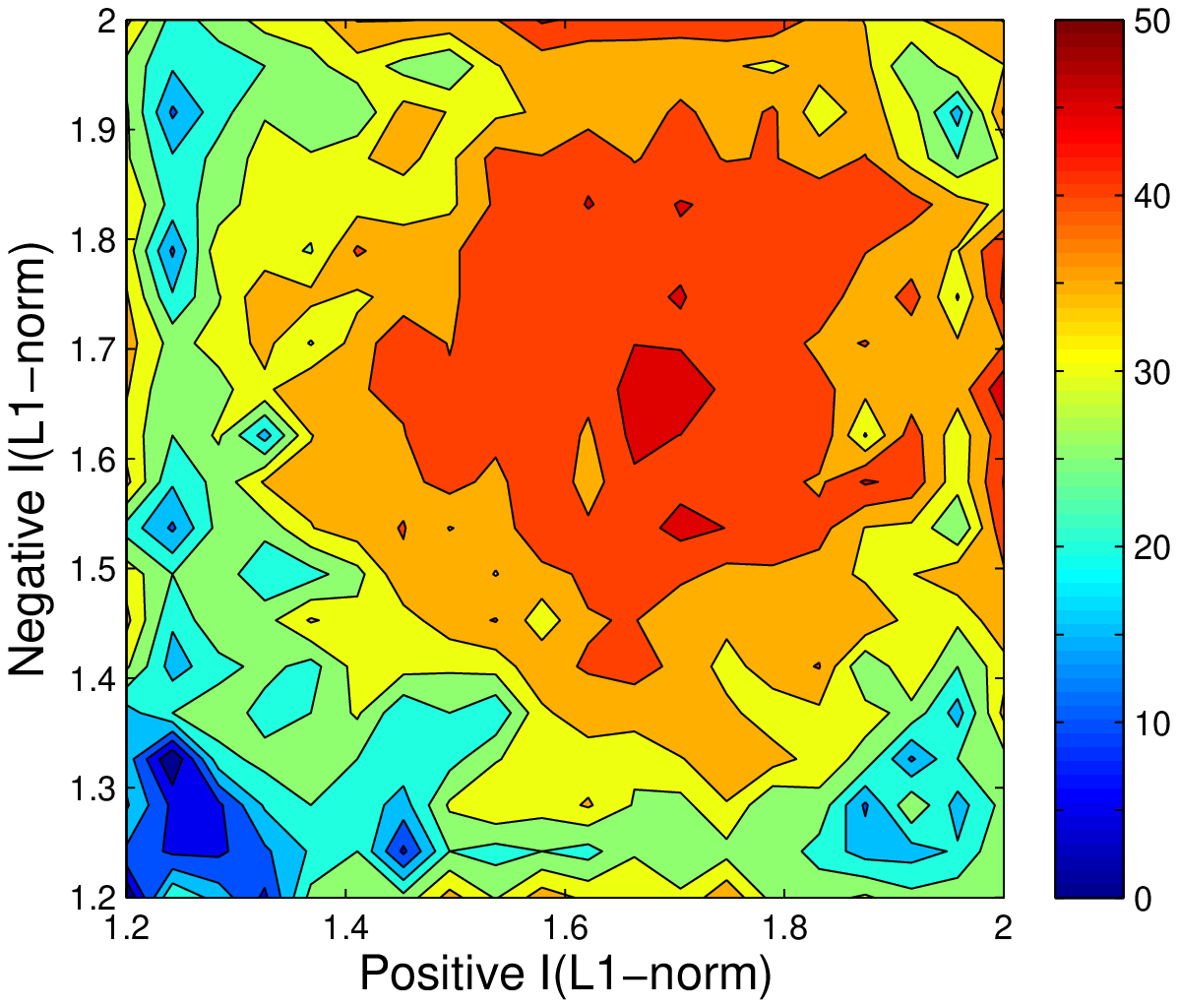}}\qquad
   \subfigure[Positive susceptibilities v.s. negative susceptibility]
   {\includegraphics[width=0.25\textwidth]{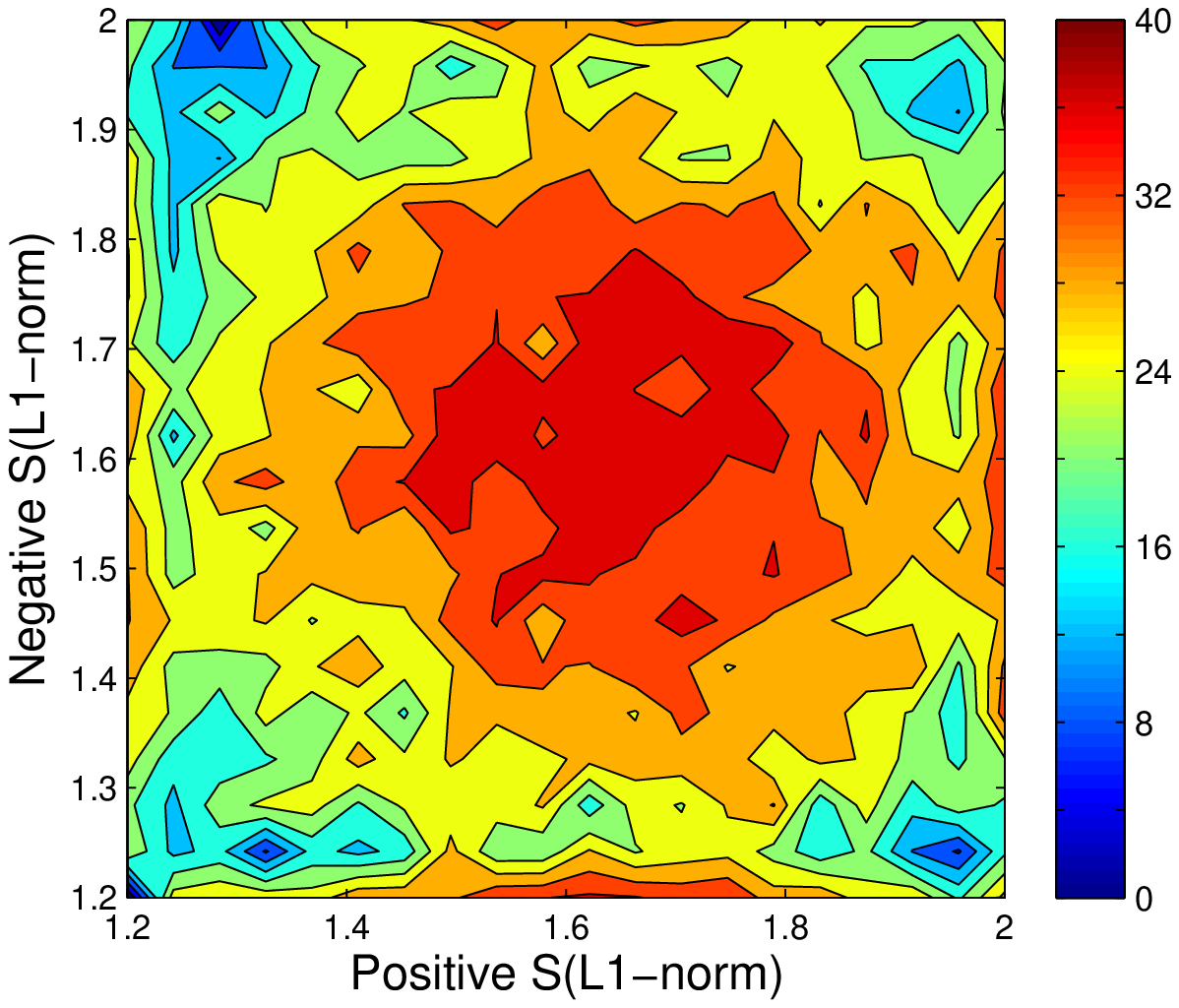}}
\caption{Analysis of L1-norm of latent sentimental influences and susceptibilities.}
\label{figIvsS}
\end{figure*}

Besides the comparisons of evaluation models,
we investigate our learned distributed representations of users on
sentiments, matrices $\mathbf I_v$ and $\mathbf S_v$ for each user $v$.
For each row in matrices $\mathbf I_v$ and $\mathbf S_v$, it
is the representation of user $v$'s influence and susceptibility on
the corresponding sentiment, denoted as
``Positive I'', ``Negative I'', ``Positive S'', and ``Negative S''. 
And we use L1-norm of those row vectors
to measure the degrees of influence and susceptibility on sentiments.
Once more, we construct points of users with those L1-norm values as
coordinates, and count the number of points falling into a predefined
lattice cell. Thus the contour maps are draw accordingly in Figure \ref{figIvsS}.
Figure \ref{figIvsS} (a) and (b) are 
the contour maps of users' influences v.s. susceptibilities
on positive sentiment and negative sentiment respectively.
There are two peaks in both contour maps. It is interesting to
see the peaks nearby ``Positive I'' and ``Negative I'' axises,
which show that amount of influential users who are not 
susceptible to others as \cite{aral2012identifying} claimed. 
We name them as \emph{original influentials} in both
positive sentiment and negative sentiment.
On the other side, there are another part of influential
users in the other two peaks located at the upper right of the contour maps,
who are susceptible and active to retweet others' messages,
named \emph{secondary influentials} in both sentiments.
In another word, the secondary influentials may take 
a lot of efforts on retweeting attractive messages to 
gain their reputations and influences. And the 
original influentials focus on composing  
attractive and initial messages for the system. 
Thus the original influentials are the primitive 
power of the system to bring new resources,
and the secondary influentials
are good advertisers to let people get information.

Finally, we show a main peak in the contour maps of Figure \ref{figIvsS} (c) and (d)
in a 2-dimensional view, 
which give a distribution of users' influences on positive
sentiment and negative sentiment in (c), and that of users' susceptibilities
on both sentiments in (d).
From Figure \ref{figIvsS} (c), it is seen that
users could have higher influences on positive sentiment, while
lower ones on negative sentiment, and vice versa, although a certain amount 
of them have almost the same high influences on both sentiments.
Figure \ref{figIvsS} (d) gives the similar solution on susceptibilities, which
some users are more sensitive to positive sentiments, and others are
sensitive to negative ones. And it seems that 
more users have the same high susceptibilities
on both sentiments than whom have the same high influences in the dataset.

\section{Conclusions}
\label{secconclusions}

We propose a model to learn the distributed representations of users' influences on 
sentiments from their history behaviors.
By explicitly characterizing the
sentimental influence and susceptibility of each user with two matrices 
respectively, the model reduces the complexity of 
pair-wise models, and in turn remits the overfitting problem.
We also design
an effective algorithm to train the model based on maximizing logarithmic
likelihood of information cascades. 
Adadelta method is used to estimate an efficient learning rate adaptively, and
PG method guarantees the constants of non-negative parameters.
Our model does not require the knowledge of
social network structure, hence having wide applicability to the scenarios with
or without explicit social networks. Explicit social network can 
be added as indicators in the likelihood of a user getting infected
by the connected and infective ones.
We evaluated the effectiveness of our model
on Microblogging dataset from Sina Weibo,
the largest social media in China. Experimental results demonstrate that our
model consistently outperforms existing pair-wise methods at predicting
cascade dynamics, ``who will be retweeted'', and cascade size prediction. 
Moreover, 
with the analysis of users' sentimental influences and susceptibilities, 
we find that there are two peaks in the contour maps, indicating 
original influentials and secondary influentials.
The former only create initial and high-quality messages to influence
others, while the latter attract others' attentions by retweeting 
interesting messages.
Besides,  
users may have different reactions on messages with different sentiments.
In the future, we would like to apply the distributed representations of
users to more imaginative applications.

\section{Acknowledgments}
This work was funded by  
National Grand Fundamental Research 973 Program of China (No. 2013CB329602, 
No. 2013CB329606), 
and the National Natural Science Foundation of China with Nos 61572467, 61232010.
The authors thank the Crowd-sourcing platform (\url{http://www.cnpameng.com/})
providing initial Sina Weibo data.

\bibliographystyle{abbrv}
\balance
\bibliography{social}

\end{document}